\documentclass[prl,showpacs,onecolumn,preprint]{revtex4}
\usepackage{amsmath}
\usepackage{amsfonts}

\begin{document}

\preprint{}
\title[Effective Simulation of Quantum Entanglement]{Effective Simulation of
Quantum Entanglement using Classical Fields Modulated with Pseudorandom
Phase Sequences}
\author{Jian Fu, Xingkun Wu}
\affiliation{State Key Lab of Modern Optical Instrumentation, Department of Optical
Engineering, Zhejiang University, Hangzhou 310027, China}

\begin{abstract}
An effective simulation of quantum entanglement is presented using classical
fields modulated with $n$ pseudorandom phase sequences (PPSs) that
constitute a $n2^{n}$-dimensional Hilbert space with a tensor product
structure. Applications to classical fields are examplied by effective
simulation of both Bell and GHZ states, and a correlation analysis was
performed to characterize the simulation. Results that strictly comply with
criteria of quantum entanglement were obtained and the approach was also
shown to be applicable to a system consisting of $n$ quantum particles.
\end{abstract}

\date{today}
\startpage{1}
\email{jianfu@zju.edu.cn}
\maketitle

Quantum entanglement, one of the most fascinating and important features in
quantum theory, is widely appreciated as an essential ingredient in quantum
computations \cite{Nielsen,Bennett,Jozsa,Einstein,Bell}. Simulations of
quantum entanglement through optical approaches were investigated both
theoretically and experimentally \cite%
{Tan,Hard,Hessmo,Cerf,Lee1,Spreeuw,Spreeuw2,Matias,Francisco,Massar,Lee2,Dragoman}%
. A quantum bit can be represented by a distinct path or space mode of a
classical field in an interferometric setup as classical optics analogies 
\cite{Cerf,Lee1,Spreeuw,Spreeuw2,Matias,Francisco,Massar,Lee2,Dragoman}.
However, a $n$-qubit system with $2^{n}$ basis states must be represented by 
$2^{n}$ distinct paths or modes of a classical field. These simulations are
usually not effective due to an exponential increase in required physical
resources correlated with the addition of quantum bits \cite{Cerf,Jozsa1}.
The drawback can be traced to a lack of a rigorous tensor-product structure
of the system \cite{Jozsa,Jozsa1,Blume}. It is very inspiring that polarized
beams (radially and azimuthally) of classical field exhibit a tensor product
structure and are isomorphic to the Bell states by adding some degrees of
freedom of a single system \cite{Toppel}.

In this letter, we present an effective simulation of quantum entanglement
of $n$ quantum bits by using an analogy of classical fields modulated with
pseudo-random phase sequences (PPSs). Based on the properties of PPSs, we
proved that the $n$ fields modulated with $n$ different PPSs constitute a $%
n2^{n}$-dimensional Hilbert space with a tensor product structure, which
differs significantly from those in classical simulations that were executed
lacking a tensor-product structure \cite%
{Cerf,Lee1,Spreeuw,Spreeuw2,Matias,Francisco,Massar,Lee2,Dragoman}. By using
an optical interferometric setup, PPSs yield not only random measurement
results, but also an ensemble model to define the ensemble average and
correlation functions \cite{Fu2}. The PPSs, derived from orthogonal
pseudorandom sequences, are widely applied to Code Division Multiple Access
(CDMA) communication technology as a way to distinguish different users \cite%
{CDMA,PS,Zigangirov}. A set of pseudorandom sequences is generated by using
a shift register guided by a Galois field $GF(p)$ that satisfies orthogonal,
closure and balance properties \cite{PS}. In this letter, we utilize a
m-sequence of period $N-1(N=p^{s})$ generated by a primitive polynomial of
degree $s$ over $GF(p)$ and apply it to binary phase shift modulation, a
well-known modulation format in wireless and optical communications \cite%
{CDMA,Zigangirov}. Next we generate a PPS set $\Xi =\left\{ \lambda ^{\left(
1\right) },\lambda ^{\left( 2\right) },\ldots \lambda ^{\left( N\right)
}\right\} $ over $GF(2)$, where each $\lambda ^{\left( i\right) }$ is a
phase sequence with $N$ phase units and time slots: $\lambda ^{\left(
i\right) }=[\lambda _{1}^{\left( i\right) },\lambda _{2}^{\left( i\right)
},\cdots \lambda _{N}^{\left( i\right) }]$, while $\lambda ^{\left( 1\right)
}$ is an all-$0$ sequence and other sequences can be generated by using
following method \cite{PS,William}: (1) given a primitive polynomial of
degree s over $GF(2)$, a base sequence of a length $2^{s}-1$ is generated by
the Linear Feedback Shift Register; (2) other sequences are obtained by
cyclic shifting of the base sequence; (3) by adding zeroes to the sequences,
the occurrence of any element equals to $2^{s}-1$; (4) mapping the elements
of the sequences to $\left[ 0,2\pi \right] $: $0$ mapping $0$, $1$ mapping $%
\pi $.

We first consider two orthogonal modes (polarization or transverse modes), $%
\left\vert 0\right\rangle $ and $\left\vert 1\right\rangle $, of a classical
field. A simulation state can be expressed as a mode superposition: $%
\left\vert \psi \right\rangle =\alpha \left\vert 0\right\rangle +\beta
\left\vert 1\right\rangle $, where $|\alpha |^{2}+|\beta |^{2}=1,(\alpha
,\beta \in 
\mathbb{C}
)$. All of the mode superposition states span a Hilbert space, where we will
explore properties associated with this special classical field. By
introducing a map $f:\lambda \rightarrow e^{i\lambda }$ on the set of $\Xi $%
, we obtain a phase sequence set $%
\Omega
=\{\varphi ^{(j)}|\varphi ^{(j)}=e^{i\lambda ^{(j)}},j=1\ldots N\}$, and
with which were written a superposed state corresponding to $n$-th sequence:%
\begin{equation}
\left\vert \psi _{n}\right\rangle \equiv e^{i\lambda ^{\left( n\right)
}}\left( \alpha _{n}\left\vert 0\right\rangle +\beta _{n}\left\vert
1\right\rangle \right)  \label{e1}
\end{equation}

According to the properties of m-sequence, the set $\Omega $ has following
properties: (a) closure: the product of any two sequences equals one of
sequence in the set; (b) balance: except $\varphi ^{(1)}$, any sequence of
set $\Omega $ satisfy $\sum_{k=1}^{N}e^{i\theta }\varphi
_{k}^{(j)}=\sum_{k=1}^{N}e^{i(\theta +\lambda _{k}^{(j)})}=0,\forall \theta
\in 
\mathbb{R}
$; (c) orthogonality: any two of the sequences satisfy normalized
correlation: $E(\varphi ^{(i)},\varphi ^{(j)})=\frac{1}{N}%
\sum_{k=1}^{N}\varphi _{k}^{(i)}\varphi _{k}^{(j)\ast }$, which equals $1$
when $i=j$, and $0$ otherwise. In fact, the map $f$ corresponds to phase
modulations of PPSs of $\Xi $ onto the classical field.

A PPS map $f$ constitutes a phase ensemble, wherein each phase unit
represents a single simulation, and measurement of a physical quantity is a
result of ensemble average. Similar to that in quantum mechanics, ensemble
average and correlation measurement can be defined \cite{CDMA,PS,Zigangirov}%
. In the quadrature demodulation, each code obtained in a sequence unit of a
PPS can be considered as a single measurement. The sequence number of the
PPS's unit can be used to label the ensemble. Different from the ergodicity
hypothesis of quantum mechanics, the ergodicity of PPS is determined and
much more efficient.

Given the properties of the PPSs and the Hilbert space, the inner product of
any two states and their orthogonal property can be obtained by:%
\begin{equation}
\left\langle \psi _{i}|\psi _{j}\right\rangle =\frac{1}{N}%
\sum_{k=1}^{N}e^{i\left( \lambda _{k}^{(j)}-\lambda _{k}^{(i)}\right)
}\left( \alpha _{i}^{\ast }\alpha _{j}+\beta _{i}^{\ast }\beta _{j}\right)
=\left\{ 
\begin{array}{c}
1,i=j \\ 
0,i\neq j%
\end{array}%
\right.   \label{e2}
\end{equation}%
where $\lambda _{k}^{(i)},\lambda _{k}^{(j)}$ are the $k$-th units of $%
\lambda ^{(i)}$ and $\lambda ^{(j)}$, respectively. Based on above
properties, the classical fields modulated with different PPSs are
independent and distinguishable. Fig. 1 shows construction pathway of
simulation states, generated by unitary transformed from initial
states------the mode superposition of classical fields with PPS $\lambda
^{(j)}$. Furthermore, a general form of a simulation state can be
constructed from $\left\vert \psi _{n}\right\rangle $:%
\begin{equation}
\left\vert \psi _{n}\right\rangle =\sum_{i=1}^{N}\alpha
_{n}^{(i)}e^{i\lambda ^{(i)}}\left\vert 0\right\rangle +\sum_{j=1}^{N}\beta
_{n}^{(j)}e^{i\lambda ^{(j)}}\left\vert 1\right\rangle   \label{e3}
\end{equation}%
Following the pathway in Fig. 1, a simulation state $\left\vert \Psi
\right\rangle $ is obtained, denoting with a direct product of $\left\vert
\psi _{n}\right\rangle $:%
\begin{equation}
\left\vert \Psi \right\rangle =\left\vert \psi _{1}\right\rangle \otimes
\left\vert \psi _{2}\right\rangle \ldots \otimes \left\vert \psi
_{N}\right\rangle   \label{e4}
\end{equation}%
Due to closure property of PPS, the phase sequence $e^{i\lambda ^{\left(
j\right) }}$ of each state $\left\vert i_{1}i_{2}\ldots i_{N}\right\rangle $
remains to belong to $\Omega $ and consists of the product of multiple
sequences. Basis for Hilbert space of simulation is spanned by $\left\{
e^{i\lambda ^{(j)}}\left\vert i_{1}i_{2}\ldots i_{N}\right\rangle |j=1\ldots
N,i_{n}=0or1\right\} $, with a total base state number of $N2^{N}$.
Generally a simulation state takes the form: 
\begin{equation}
\left\vert \Psi \right\rangle =\sum_{i_{1}=0}^{1}\ldots \sum_{i_{N}=0}^{1}
\left[ \sum_{j=1}^{N}C_{i_{1}\ldots i_{N}}^{(j)}e^{i\lambda ^{\left(
j\right) }}\left\vert i_{1}i_{2}\ldots i_{N}\right\rangle \right] 
\label{e5}
\end{equation}%
where $C_{i_{1}\ldots i_{N}}^{(j)}$ denotes a total of $N2^{N}$
coefficients. It is obvious that the Hilbert simulation space is greater
than what is required for simulation of quantum state. To obtain a space the
same size as that in quantum mechanics, either restrictions or a proper
measurement need to apply \cite{Fu2}.

PPS provides not only the tensor structure and space needed for quantum
state simulation, it also yields the property that an entangled state cannot
be expressed in terms of direct product of tensors by using PPS properties
and phase ensemble average. In the following we use density matrix to
illustrate this feature. We assume that a simple type of simulation state of 
$N$ fields can be expressed: 
\begin{equation}
\left\vert \Psi \right\rangle =\left\vert \psi _{1}\right\rangle \otimes
\left\vert \psi _{2}\right\rangle \ldots \otimes \left\vert \psi
_{N}\right\rangle =e^{i\lambda ^{sum}}\left( \sum_{i=1}^{N^{\prime
}}C_{i}\left\vert x_{i}\right\rangle +\sum_{j=1}^{N^{^{\prime \prime
}}}C_{j}e^{i\lambda ^{(j)}}\left\vert x_{j}\right\rangle \right)  \label{e6}
\end{equation}%
where $N^{\prime }+N^{^{\prime \prime }}=2^{N}$ and $N^{^{\prime \prime }}<N$%
, $\lambda ^{sum}=\sum_{n=1}^{N}\lambda ^{(n)}$, and $\left\vert
x_{i,j}\right\rangle =\left\vert i_{1}i_{2}\ldots i_{N}\right\rangle $. A
density matrix $\rho \ $can be calculated:%
\begin{equation}
\rho \equiv \left\vert \Psi \right\rangle \left\langle \Psi \right\vert
=e^{i\lambda ^{sum}}\left( \sum_{i=1}^{N^{\prime }}C_{i}\left\vert
x_{i}\right\rangle +\sum_{j=1}^{N^{^{\prime \prime }}}C_{j}e^{i\lambda
^{(j)}}\left\vert x_{j}\right\rangle \right) \times e^{-i\lambda
^{sum}}\left( \sum_{i=1}^{N^{\prime }}C_{i}^{\ast }\left\langle
x_{i}\right\vert +\sum_{j=1}^{N^{^{\prime \prime }}}C_{j}^{\ast
}e^{-i\lambda ^{(j)}}\left\langle x_{j}\right\vert \right)  \label{e7}
\end{equation}%
which is simplified into%
\begin{eqnarray}
\rho &=&\sum_{n=1}^{2^{N}}|C_{n}|^{2}\left\vert x_{n}\right\rangle
\left\langle x_{n}\right\vert +\sum_{i\neq i^{\prime }=1}^{N^{^{\prime
}}}\left( C_{i^{^{\prime }}}^{\ast }C_{i}\left\vert x_{i}\right\rangle
\left\langle x_{i^{^{\prime }}}\right\vert +C_{i}^{\ast }C_{i^{\prime
}}\left\vert x_{i^{^{\prime }}}\right\rangle \left\langle x_{i}\right\vert
\right)  \label{e8} \\
&&+\sum_{j\neq j^{\prime }=1}^{N^{^{\prime \prime }}}\left( C_{j^{^{\prime
}}}^{\ast }C_{j}e^{i\lambda ^{\left( l\right) }}\left\vert
x_{j}\right\rangle \left\langle x_{j^{^{\prime }}}\right\vert +C_{j}^{\ast
}C_{j^{\prime }}e^{-i\lambda ^{\left( l\right) }}\left\vert x_{j^{^{\prime
}}}\right\rangle \left\langle x_{j}\right\vert \right)  \notag \\
&&+\sum_{i=1}^{N^{^{\prime }}}\sum_{j=1}^{N^{^{\prime \prime }}}\left(
C_{i}^{\ast }C_{j}e^{i\lambda ^{\left( j\right) }}\left\vert
x_{j}\right\rangle \left\langle x_{i}\right\vert +C_{j}^{\ast
}C_{i}e^{-i\lambda ^{\left( j\right) }}\left\vert x_{i}\right\rangle
\left\langle x_{j}\right\vert \right)  \notag
\end{eqnarray}%
where $\lambda ^{(l)}=\lambda ^{(j)}-\lambda ^{\left( j^{\prime }\right) }$.
By applying phase ensemble averaging \cite{PS}, mean reduced density matrix
is defined $\tilde{\rho}\equiv \frac{1}{N}\sum_{k=1}^{N}\rho $. Due to the
balance property of PPS, then we obtain%
\begin{equation}
\tilde{\rho}=\sum_{n=1}^{2^{N}}|C_{n}|^{2}\left\vert x_{n}\right\rangle
\left\langle x_{n}\right\vert +\sum_{i\neq i^{\prime }=1}^{N^{^{\prime
}}}\left( C_{i^{^{\prime }}}^{\ast }C_{i}\left\vert x_{i}\right\rangle
\left\langle x_{i^{^{\prime }}}\right\vert +C_{i}^{\ast }C_{i^{\prime
}}\left\vert x_{i^{^{\prime }}}\right\rangle \left\langle x_{i}\right\vert
\right)  \label{e9}
\end{equation}%
Eq. (\ref{e9}) shows that all non-diagonal terms including $\left\vert
x_{j}\right\rangle $ disappear and the reduced density matrix $\tilde{\rho}$
might not be expressed in terms of a direct product of the states $%
\left\vert x_{n}\right\rangle $, similar to the case of quantum entanglement
states.

In addition to the fact that a quantum entanglement cannot be expressed in
terms of direct tensor product, quantum entanglement also make a correlation
measurement different. The correlation analysis on the simulation states is
necessary because the nonlocal correlation with Bell's inequality and
equality criterion is the most fundamental property of quantum entanglement.
In order to perform the correlation analysis, a correlation measurement $%
\hat{P}$ on $\left\vert \psi \right\rangle $ is given%
\begin{equation}
\bar{P}(\theta )=\left\langle \psi \right\vert \hat{P}(\theta )\left\vert
\psi \right\rangle =\left( 
\begin{array}{cc}
\alpha ^{\ast } & \beta ^{\ast }%
\end{array}%
\right) \left( 
\begin{array}{cc}
0 & e^{i\theta } \\ 
e^{-i\theta } & 0%
\end{array}%
\right) \binom{\alpha }{\beta }=\alpha ^{\ast }\beta e^{i\theta }+\alpha
\beta ^{\ast }e^{-i\theta }  \label{e10}
\end{equation}%
For convenience, coefficients $\alpha ,\beta $ are set to be $1/\sqrt{2}$,
yielding $\bar{P}(\theta )=\cos (\theta )$. Further we generalize $\hat{P}$
to the case of $N$ fields:%
\begin{equation}
\hat{P}(\theta _{1},\ldots \theta _{N})=\hat{P}\left( \theta _{1}\right)
\otimes \hat{P}\left( \theta _{2}\right) \otimes \ldots \hat{P}\left( \theta
_{N}\right)  \label{e11}
\end{equation}%
Then we obtain the correlation analysis of the simulation states using $\hat{%
P}$ and the density matrix $\rho $:%
\begin{eqnarray}
E\left( \theta _{1},\ldots \theta _{N}\right) &=&\frac{1}{N}\sum_{k=1}^{N}Tr%
\left[ \rho \hat{P}(\theta _{1},\ldots \theta _{N})\right] =Tr\left[ \tilde{%
\rho}\hat{P}(\theta _{1},\ldots \theta _{N})\right]  \label{e12} \\
&=&\sum_{n=1}^{2^{N}}|C_{n}|^{2}\left\langle x_{n}\right\vert \hat{P}%
\left\vert x_{n}\right\rangle +\sum_{i\neq i^{\prime }=1}^{N^{\prime
}}\left( C_{i^{^{\prime }}}^{\ast }C_{i}\left\langle x_{i^{^{\prime
}}}\right\vert \hat{P}\left\vert x_{i}\right\rangle +C_{i}^{\ast
}C_{i^{\prime }}\left\langle x_{i}\right\vert \hat{P}\left\vert
x_{i^{^{\prime }}}\right\rangle \right)  \notag
\end{eqnarray}%
Eq. (\ref{e12}) shows that only non-diagonal terms $\sum_{i\neq i^{\prime
}=1}^{N^{\prime }}\left( C_{i^{^{\prime }}}^{\ast }C_{i}\left\langle
x_{i^{^{\prime }}}\right\vert \hat{P}\left\vert x_{i}\right\rangle
+C_{i}^{\ast }C_{i^{\prime }}\left\langle x_{i}\right\vert \hat{P}\left\vert
x_{i^{^{\prime }}}\right\rangle \right) $ remain.

Key to an effective simulation of quantum entanglement is that the physical
resources for the simulation does not increase exponentially with number of
particle. In the following we discuss analysis of computation complexity. A
simple unitary transformation, $NOT$ gate, is used as an example to show
computation complexity. Starting with a single field $\left\vert \psi
_{n}\right\rangle =e^{i\lambda ^{(n)}}\left( \alpha _{n}\left\vert
0\right\rangle +\beta _{n}\left\vert 1\right\rangle \right) $, applying a
unitary transformation switching $\hat{U}:\left\vert 0\right\rangle
\leftrightarrow \left\vert 1\right\rangle $ to decomposes PPS into each
phase unit: $\hat{U}\left\vert \psi _{n}\right\rangle \rightarrow \left[
e^{i\lambda _{k}^{(n)}}\hat{U}\left( \alpha _{n}\left\vert 0\right\rangle
+\beta _{n}\left\vert 1\right\rangle \right) |k=1\ldots N\right]
=e^{i\lambda ^{(n)}}\left( \alpha _{n}\left\vert 1\right\rangle +\beta
_{n}\left\vert 0\right\rangle \right) $. For each phase unit, its
computation is the same as that in quantum computation, therefore
computation for $N$ phase units equals $N$ times of quantum computation of
each phase unit. We can extand unitary transformations to simulation states
with $N$ fields: 
\begin{eqnarray}
\hat{U} &:&\left\vert \Psi \right\rangle \rightarrow \left\vert \Psi
^{\prime }\right\rangle  \label{e13} \\
\left\vert \Psi \right\rangle &=&\sum_{i_{1}=0}^{1}\ldots \sum_{i_{N}=0}^{1} 
\left[ \sum_{j=1}^{N}C_{i_{1}\cdots i_{N}}^{(j)}e^{i\lambda
^{(j)}}\left\vert i_{1}\ldots i_{k}\ldots i_{N}\right\rangle \right]  \notag
\\
\left\vert \Psi ^{^{\prime }}\right\rangle &=&\sum_{i_{1}=0}^{1}\ldots
\sum_{i_{N}=0}^{1}\left[ \sum_{j=1}^{N}C_{i_{1}\cdots i_{N}}^{(j)^{\prime
}}e^{i\lambda ^{(j)}}\left\vert i_{1}\ldots i_{k}^{^{\prime }}\ldots
i_{N}\right\rangle \right]  \notag
\end{eqnarray}%
and coefficients $C_{i_{1}\cdots i_{N}}^{(j)}$ and $C_{i_{1}\cdots
i_{N}}^{(j)^{\prime }}$ are related by an unitary transformation:%
\begin{equation}
C_{i_{1}\cdots i_{k}^{^{\prime }}\cdots i_{N}}^{(j)^{\prime
}}=\sum_{i_{k}}U_{i_{k}^{^{\prime }}}^{i_{k}}C_{i_{1}\cdots i_{k}\cdots
i_{N}}^{(j)}  \label{e14}
\end{equation}%
Because a PPS contains $N$ phase units $e^{i\lambda _{k}^{\left( j\right) }}$
and $N$ time slots, therefore the required computation is $N$ times that of
quantum computation, but $2^{N}$ times is unnecessary \cite{Jozsa}.

\textit{Two-particles Bell states}: Consider the case that the modes $%
\left\vert 1\right\rangle $ in the states $\left\vert \psi _{a}\right\rangle 
$ and $\left\vert \psi _{b}\right\rangle $ similar to Eq. (\ref{e1}) are
exchanged by a mode exchanger constituted by mode splitters and combiners 
\cite{Fu,Fu2}. The exchange yields the following states:%
\begin{eqnarray}
\left\vert \psi _{a}^{^{\prime }}\right\rangle &=&\frac{e^{i\lambda ^{(a)}}}{%
\sqrt{2}}\left( \left\vert 0\right\rangle +e^{i\gamma ^{(a)}}\left\vert
1\right\rangle \right)  \label{e15} \\
\left\vert \psi _{b}^{^{\prime }}\right\rangle &=&\frac{e^{i\lambda ^{(b)}}}{%
\sqrt{2}}\left( \left\vert 0\right\rangle +e^{i\gamma ^{(b)}}\left\vert
1\right\rangle \right)  \notag
\end{eqnarray}%
where the relative phase sequences (RPSs) $\gamma ^{(a)}=-\gamma
^{(b)}=\lambda ^{(b)}-\lambda ^{(a)}$, and $\gamma ^{(a)}+\gamma ^{(b)}=0$.
The simulation state $\left\vert \Psi \right\rangle $ is obtained:%
\begin{equation}
\left\vert \Psi \right\rangle =\left\vert \psi _{a}^{^{\prime
}}\right\rangle \otimes \left\vert \psi _{b}^{^{\prime }}\right\rangle =%
\frac{e^{i\left( \lambda ^{(a)}+\lambda ^{(b)}\right) }}{2}\left[ \left\vert
0\right\rangle \left\vert 0\right\rangle +\left\vert 1\right\rangle
\left\vert 1\right\rangle +e^{i\gamma ^{(a)}}\left\vert 1\right\rangle
\left\vert 0\right\rangle +e^{i\gamma ^{(b)}}\left\vert 0\right\rangle
\left\vert 1\right\rangle \right]  \label{e15-1}
\end{equation}%
Appearently the reduced density matrix $\tilde{\rho}$ cannot be direct
product decomposited due to only non-diagonal term $\left\vert
00\right\rangle \left\langle 11\right\vert +\left\vert 11\right\rangle
\left\langle 00\right\vert $ remaining.

Then we obtain the results of the fields in the correlation measurement $%
\bar{P}(\theta _{a},k)=\cos (\theta _{a}+\gamma _{k}^{(a)})$ and $\bar{P}%
(\theta _{b},k)=\cos (\theta _{b}+\gamma _{k}^{(b)})$, where $\gamma
_{k}^{(a)},\gamma _{k}^{(b)}$ are the $k$-th units of the RPSs $\gamma
^{(a)} $ and $\gamma ^{(b)}$, respectively. Then the correlation function is%
\begin{equation}
E(\theta _{a},\theta _{b})=\frac{1}{NC}\sum_{k=1}^{N}\bar{P}(\theta _{a},k)%
\bar{P}(\theta _{b},k)=\cos (\theta _{a}+\theta _{b})  \label{e16}
\end{equation}%
where $C=1/2$ is the normalization coefficient. The states in Eq. (\ref{e15}%
) are considered to be a classical field simulation of the Bell state $%
\left\vert \Psi ^{+}\right\rangle $. By substituting the above correlation
functions into Bell inequality (CHSH inequality) \cite{CHSH}:%
\begin{equation}
\left\vert B\right\vert =\left\vert E(\theta _{a},\theta _{b})-E(\theta
_{a},\theta _{b}^{\prime })+E(\theta _{a}^{\prime },\theta _{b}^{\prime
})+E(\theta _{a}^{\prime },\theta _{b})\right\vert =2\sqrt{2}>2  \label{e17}
\end{equation}%
where $\theta _{a},\theta _{a}^{\prime },\theta _{b}$ and $\theta
_{b}^{\prime }$ are $\pi /4,-\pi /4,0$ and $\pi /2$, respectively, when
Bell's inequality is maximally violated.

Bell state $\left\vert \Psi ^{+}\right\rangle $ differs from $\left\vert
\Psi ^{-}\right\rangle $ by $\pi $ phase. Similarly, simulation of the Bell
state $\left\vert \Psi ^{-}\right\rangle $ is expressed as $\left\vert \psi
_{a}^{^{\prime }}\right\rangle =e^{i\lambda ^{(a)}}\left( \left\vert
0\right\rangle +e^{i\gamma ^{(a)}}\left\vert 1\right\rangle \right) /\sqrt{2}%
,\left\vert \psi _{b}^{^{\prime }}\right\rangle =e^{i\lambda ^{(b)}}\left(
\left\vert 0\right\rangle +e^{i\left( \gamma ^{(b)}+\pi \right) }\left\vert
1\right\rangle \right) /\sqrt{2}$. By performing the transformation $\hat{%
\sigma}_{x}:\left\vert 0\right\rangle \leftrightarrow \left\vert
1\right\rangle $ on $\left\vert \psi _{b}^{^{\prime }}\right\rangle $ of the
state $\left\vert \Psi ^{\pm }\right\rangle $, we obtain the simulation of
the Bell state $\left\vert \Phi ^{+}\right\rangle $ expressed as $\left\vert
\psi _{a}^{^{\prime }}\right\rangle =e^{i\lambda ^{(a)}}\left( \left\vert
0\right\rangle +e^{i\gamma ^{(a)}}\left\vert 1\right\rangle \right) /\sqrt{2}%
,\left\vert \psi _{b}^{^{\prime }}\right\rangle =e^{i\lambda ^{(b)}}\left(
\left\vert 1\right\rangle +e^{i\gamma ^{(b)}}\left\vert 0\right\rangle
\right) /\sqrt{2}$, and of $\left\vert \Phi ^{-}\right\rangle $ expressed as 
$\left\vert \psi _{a}^{^{\prime }}\right\rangle =e^{i\lambda ^{(a)}}\left(
\left\vert 0\right\rangle +e^{i\gamma ^{(a)}}\left\vert 1\right\rangle
\right) /\sqrt{2},\left\vert \psi _{b}^{^{\prime }}\right\rangle
=e^{i\lambda ^{(b)}}\left( \left\vert 1\right\rangle +e^{i\left( \gamma
^{(b)}+\pi \right) }\left\vert 0\right\rangle \right) /\sqrt{2}$. Then their
correlation functions $E_{\Psi ^{-}}\left( \theta _{a},\theta _{b}\right)
=-\cos \left( \theta _{a}+\theta _{b}\right) ,E_{\Phi ^{\pm }}\left( \theta
_{a},\theta _{b}\right) =\pm \cos \left( \theta _{a}-\theta _{b}\right) $
are obtained. To substitute the correlation functions into Eq. (\ref{e17}),
we also obtain the maximal violation of Bell's inequality. The violation of
Bell's criterion demonstrates the nonlocal correlation of the two classical
fields in our simulation, which results from shared randomness of the PPSs.

\textit{GHZ states}: The nonlocality of the multipartite entangled GHZ
states can in principle be manifest in a new criterion and need not be
statistical as the violation of Bell inequality \cite{GHZ}. Preparing three
states $\left\vert \psi _{a}\right\rangle ,\left\vert \psi _{b}\right\rangle 
$ and $\left\vert \psi _{c}\right\rangle $ similar to Eq. (\ref{e1}), and by
cyclically exchanging the modes $\left\vert 1\right\rangle $ of the states,
we obtain the states as following%
\begin{eqnarray}
\left\vert \psi _{a}^{^{\prime }}\right\rangle &=&\frac{e^{i\lambda ^{(a)}}}{%
\sqrt{2}}\left( \left\vert 0\right\rangle +e^{i\gamma ^{(a)}}\left\vert
1\right\rangle \right)  \label{e18} \\
\left\vert \psi _{b}^{^{\prime }}\right\rangle &=&\frac{e^{i\lambda ^{(b)}}}{%
\sqrt{2}}\left( \left\vert 0\right\rangle +e^{i\gamma ^{(b)}}\left\vert
1\right\rangle \right)  \notag \\
\left\vert \psi _{c}^{^{\prime }}\right\rangle &=&\frac{e^{i\lambda ^{(c)}}}{%
\sqrt{2}}\left( \left\vert 0\right\rangle +e^{i\gamma ^{(c)}}\left\vert
1\right\rangle \right)  \notag
\end{eqnarray}%
where the RPSs $\gamma ^{(a)}=\lambda ^{(b)}-\lambda ^{(a)},\gamma
^{(b)}=\lambda ^{(c)}-\lambda ^{(b)},\gamma ^{(c)}=\lambda ^{(a)}-\lambda
^{(c)}$ and $\gamma ^{(a)}+\gamma ^{(b)}+\gamma ^{(c)}=0$. We obtain the
measurement results $\bar{P}(\theta _{a},k)=\cos (\theta _{a}+\gamma
_{k}^{(a)}),\bar{P}(\theta _{b},k)=\cos (\theta _{b}+\gamma _{k}^{(b)}),\bar{%
P}(\theta _{c},k)=\cos (\theta _{c}+\gamma _{k}^{(c)})$ for the states $%
\left\vert \psi _{a}^{^{\prime }}\right\rangle ,\left\vert \psi
_{b}^{^{\prime }}\right\rangle $ and $\left\vert \psi _{c}^{^{\prime
}}\right\rangle $ in the correlation measurement, respectively, and the
correlation function%
\begin{equation}
E(\theta _{a},\theta _{b},\theta _{c})=\frac{1}{NC}\sum\limits_{k=1}^{N}\bar{%
P}(\theta _{a},k)\bar{P}(\theta _{b},k)\bar{P}(\theta _{c},k)=\cos (\theta
_{a}+\theta _{b}+\theta _{c})  \label{e19}
\end{equation}%
where $C=1/4$ is the normalized coefficient. If $\theta _{a}+\theta
_{b}+\theta _{c}=0,E(\theta _{a},\theta _{b},\theta _{c})=1$. If $\theta
_{a}+\theta _{b}+\theta _{c}=\pi ,E(\theta _{a},\theta _{b},\theta _{c})=-1$%
. By using GHZ State, the family of simple proofs of Bell's theorem without
inequalities can be obtained [26], which is different from the criterion of
CHSH inequality. The sign of the correlation function can be also treated as
the criterion, such as the negative correlation for nonlocal and the
positive correlation for local when $\theta _{a}=\pi /3,\theta _{b}=\pi
/3,\theta _{c}=\pi /3$. We obtain that the simulation state in Eq. (\ref{e18}%
) shows the negative correlation. The results are similar to the quantum
case of GHZ states.

Further, the simulation of GHZ state could be generalized to the case of $N$
particles. By preparing $N$ states similar to Eq. (\ref{e1}) and cyclically
exchanging the modes $\left\vert 1\right\rangle $ of the states, the RPSs
satisfy $\gamma ^{(1)}+\cdots +\gamma ^{(N)}=0$. We obtain the correlation
function%
\begin{equation}
E(\theta _{1},\ldots \theta _{N})=\frac{1}{NC}\sum\limits_{k=1}^{N}\bar{P}%
(\theta _{1},k)\ldots \bar{P}(\theta _{N},k)=\cos \left( \theta _{1}+\cdots
+\theta _{N}\right)  \label{e20}
\end{equation}%
where $\bar{P}(\theta _{i},k)=\cos (\theta _{i}+\gamma _{k}^{(i)})$ is the
result of the classical field with $i$-th RPSs at the $k$-th sequence units
in the correlation measurement, and $C=1/2^{N-1}$ is the normalized
coefficient.

Using the same notion, we can obtain simulation results of other quantum
entanglement states. It should be pointed out that the phase randomness
provided by PPSs is different from the case of quantum mixed states. Quantum
mixed states result from decoherence and all coherent superposition items
disappear. In contract to the decoherence, some coherent superposition items
remain in the simulation state due to the constraints of the RPSs, such as $%
\gamma ^{(a)}+\gamma ^{(b)}=0,\gamma ^{(a)}+\gamma ^{(b)}+\gamma ^{(c)}=0$
for the simulation of Bell states and GHZ state, respectively. These
remaining items make it possible to simulate quantum entangled pure states.

As shown in the above examples, we utilized the properties of PPSs to label
classical fields that are even overlapped in the same space and time. In
simulation of entangled states, the resources required are the PPSs instead
of classical field modes. It means that the amount of PPSs grows linearly
with the number of quantum particles. According to the m-sequence theory,
the number of PPSs in the set equals to the length of sequences, which means
that the time resource (the length of sequence) required also grows linearly
with the number of the particles.

In conclusion, a novel simulation method for quantum entanglement is
presented, with its mathematical expressions and physical meanings identical
to those in quantum mechanics. In the framework of quantum mechanics, the
overall phase of a wavefunction can be ignored, as it has no contribution to
the probability distribution. However, quantum entanglement must be related
to two or more spatially separable quantum particles. By introducing a phase
factor to superposed states with PPS properties, we conclude that quantum
entanglement can be efficiently simulated by using a classical field
modulated with PPSs. The research on this simulation not only provides
useful insights into fundamental features of quantum entanglement, but also
yields new insights into quantum computation.

Acknowledgement: \textit{Supported by the National Natural Science
Foundation of China under Grant No 60407003 and 61178049.}

\begin{description}
\item \newpage

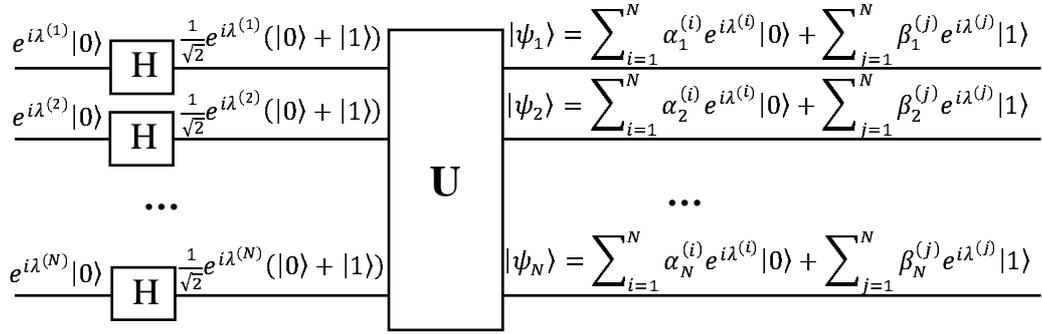
\item[Fig.1] Construction pathway of simulation states is displayed, which
are generated by unitary transformed from initial states------the mode
superposition of classical fields with PPS $\lambda ^{(i)}$.
\end{description}

\end{document}